\documentclass[aip,graphicx,reprint]{revtex4-1}

\usepackage{graphicx}
\usepackage{dcolumn}
\usepackage{bm}
\draft 

\begin{document}


\title{Hanle detection for optical clocks} 



\author{Xiaogang Zhang}
\author{Shengnan Zhang}
\author{Duo Pan}
\author{Peipei Chen}
\author{Xiaobo Xue}
\author{Wei Zhuang}

\author{Jingbiao Chen}

\email{jbchen@pku.edu.cn}

\affiliation{State Key Laboratory of Advanced Optical Communication Systems and Networks, Institute of Quantum Electronics, School of Electronics Engineering $\&$ Computer Science, Peking University, Beijing 100871, China}


\date{\today}

\begin{abstract}
Considering the strong inhomogeneous spatial polarization and intensity distribution of spontaneous decay fluorescence due to the Hanle effect, we propose and demonstrate a universe Hanle detection configuration of electron-shelving method for optical clocks. Experimental results from Ca atomic beam optical frequency standard with 423 nm electron-shelving method show that a designed Hanle detection geometry with optimized magnetic field direction, detection laser beam propagation and polarization direction, and detector position can improve the fluorescence collection rate by more than one order of magnitude comparing with that of inefficient geometry. With the fixed 423 nm fluorescence, the improved 657 nm optical frequency standard signal intensity is presented. And the potential application of the Hanle detection geometry designed for facilitating the fluorescence collection for optical lattice clock with a limited solid angle of the fluorescence collection has been discussed. This Hanle detection configuration is also effective for ion detection in ion optical clock and quantum information experiments. Besides, a cylinder fluorescence collection structure is designed to increase the solid angle of the fluorescence collection in Ca atomic beam optical frequency standard.
\end{abstract}


\maketitle 


The Hanle effect, dating back to 1923\cite{Hanle1923}, has contributed much in the development of atomic physics and, more generally, of quantum mechanics. Hanle had first correctly elaborated the effect named after him in 1924\cite{Hanle1924}. Hanle effect is mainly known as a depolarization of resonance fluorescence as the Zeeman states of the excited atomic energy levels become degenerate when an applied magnetic field is swept through zero. The Hanle effect has been used to precisely measure the lifetimes of excited atomic states\cite{lifetime1, lifetime2, lifetime3}. Its extension to nonzero magnetic fields, the level-crossing technique, made possible precise sub-Doppler measurements of fine and hyperfine structure separation\cite{Hanle, atomic5}. It plays an irreplaceable role in modern atomic spectroscopy\cite{atomic5, atomic1, atomic2, Yu, atomic6, atomic3, Co}. Pierre C$\acute{e}$rez et.al. found that the optical pumping efficiency as well as the fluorescence yield are sensitive to the Hanle effect in the Cesium beam frequency standard\cite{Cs1, Cs2}.

 \begin{figure}
\includegraphics[width=6cm]{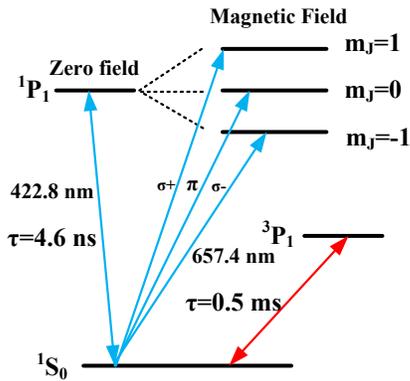}
\caption{\label{fig:epsart} Relevant Ca energy levels.}
\end{figure}

In our proposed Ca atomic beam optical frequency standard\cite{Ca-p} with the electron-shelving method, we use 423 nm $^{1}S_{0}$ $\rightarrow$ $^{1}P_{1}$ transition laser to obtain the 657 nm $^{1}S_{0}$ $\rightarrow$ $^{3}P_{1}$ clock transition signal, which is the frequency standard. Because of the fast cycling transition of the 423 nm transition, we collect the enhanced fluorescence of the 423 nm transition to improve the signal to noise ratio of the 657 nm frequency standard signal. However, the strong influence of the Hanle effect on the fluorescence collection rate of the electron-shelving method, which is the most common detection method in current neutral and ion optical clocks\cite{Lattice1, Ye, Lattice3, Ch, Gao}, has never been discussed in Ca atomic beam optical frequency standard\cite{Ca-p, Ca-a}, or the other optical lattice clocks\cite{Lattice1, Ye, Lattice3, Ch}. In this paper, it is demonstrated that the Hanle effect strongly affects the detection efficiency of electron-shelving method in a compact Ca atomic beam optical frequency standard. We propose and demonstrate a universe Hanle detection configuration of the electron-shelving method for optical clocks. The Hanle detection geometry with optimized magnetic field direction, detection laser beam propagation and polarization direction, and detector position can improve the detected fluorescence intensity by more than one order of magnitude comparing with that of inefficient geometry in optical clocks. The improved 657 nm optical frequency standard signal is presented with the optimized Hanle detection configuration. Besides, we design a fluorescence collection structure to improve the solid angle of the fluorescence collection in Ca atomic beam optical frequency standard.

Nowadays, the optical lattice clock has achieved $10^{-18}$ stability and uncertainty\cite{Lattice1, Ye}, which is the best performance of optical clocks. With the optimized Hanle detection geometry, a high signal to noise ratio of the frequency standard signal can be achieved with the enhanced fluorescence intensity. This will be very helpful to further improve the stability of optical clocks\cite{Lattice1, Ye, Lattice3, Ch}. This Hanle detection configuration is also effective for ion detection in ion optical clock and quantum information experiments\cite{Gao, Qu, Quantum}.

 In Ca atomic beam optical frequency standard\cite{Ca-p}, we collect the 423 nm transition fluorescence signal to steer the 657 nm frequency standard signal. The relevant energy levels are shown in Figure 1. With the presence of the magnetic field, the excited state $^{1}P_{1}$ is split into three Zeeman sub-levels. The whole 423 nm fluorescence collection system is shown in Figure 2. In Figure 2, the orientation of the homogeneous magnetic field is in the $\hat{x}-$direction. The Ca atomic beam is along the $\hat{y}-$direction. The incident laser travels along the $\hat{z}-$direction and a linear polarization orientation can be changed in the $\hat{x}-\hat{y}$ plane. After the interaction between Ca atomic beam and 423 nm laser, the photomultiplier(PMT) collects the fluorescence radiation along the $\hat{x}-$direction. Besides, the polarizer in Figure 2 is only used for investigating the Hanle effect and absent for the fluorescence collection in optical frequency standard.

 In our preliminary fluorescence collection system, the cylinder fluorescence collection structure with two through-holes in the dashed box as shown in Figure 2 is absent. Without the cylinder fluorescence collection structure, the detected fluorescence intensity of PMT changed as the linear polarization orientation of 423 nm laser changed without the polarizer. The maximum fluorescence intensity occurred with the $\hat{y}-$direction polarization of the detection laser while the minimum intensity occurred with the $\hat{x}-$direction polarization of the detection laser. The detected fluorescence intensity of PMT changing with 423 nm laser polarization near zero static magnetic field is shown in Figure 3. In Figure 3, the detected fluorescence intensity with the $\hat{y}-$direction polarization of the detection laser is 15.8 times larger than the detected fluorescence intensity with the $\hat{x}-$direction polarization of the detection laser.

\begin{figure}
\includegraphics[width=8cm]{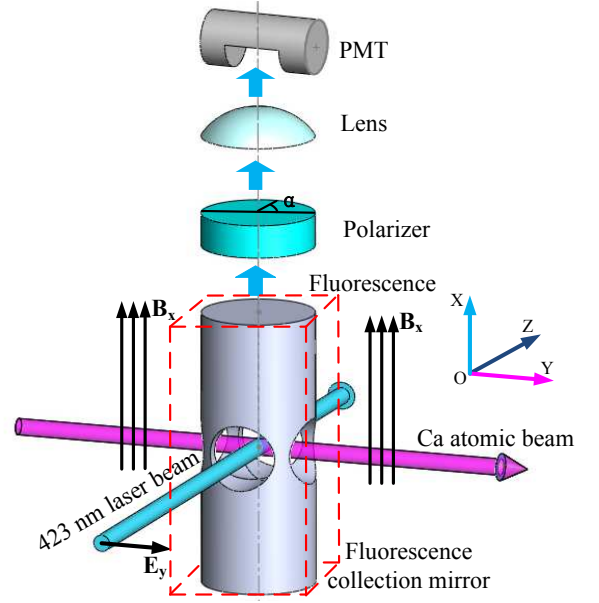}
\caption{\label{fig:epsart} The Hanle detection configuration. The cylinder fluorescence collection structure with two through-holes in the interaction area in the dashed box is added for increasing the solid angle of the fluorescence collection. Without the cylinder fluorescence collection structure, the Hanle effect is measured.}
\end{figure}

\begin{figure}
\centering\includegraphics[width=9cm]{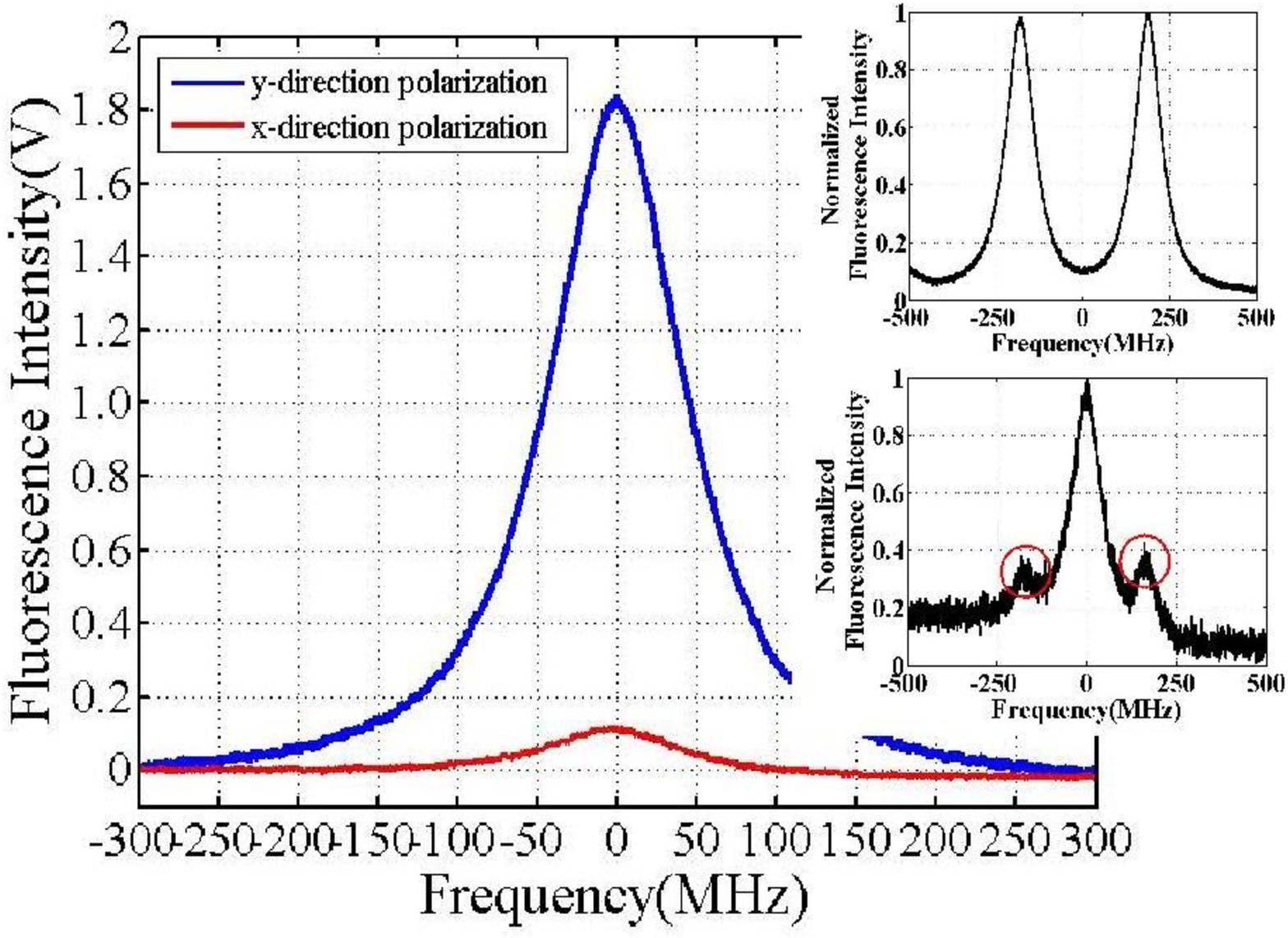}
\caption{\label{fig:epsart} The detected fluorescence signal in different polarization orientation of the 423 nm detection laser with zero magnetic field. The upper blue fluorescence signal is collected when the polarization orientation of the 423 nm laser is in the $\hat{y}-$direction; The lower red fluorescence signal is collected when the polarization orientation of the 423 nm laser is in the $\hat{x}-$direction; The upper inset is the detected fluorescence signal as the polarization orientation of the 423 nm laser is in the $\hat{y}-$direction while the magnetic field is 121.6 Gauss. The lower inset is the detected fluorescence signal as the polarization orientation of the 423 nm laser is in the $\hat{x}-$direction while the magnetic field is 121.6 Gauss.}
\end{figure}

With fixed magnetic field orientation and fixed laser propagation direction, the detected fluorescence intensity under zero magnetic field changes with the laser polarization orientation. According to the classical and quantum theory, the intensity of the fluorescence can be expressed as\cite{Hanle}

\begin{eqnarray}
 I(B)&=&C\int^{\infty}_{0}I(B,t)dt\nonumber\\
      &=&\frac{CI_{0}}{2}\int^{\infty}_{0}\exp{\{-\Gamma t\}}[1-\cos2(w_{L}t-\alpha)]dt \nonumber\\
      &=&\frac{I_{0}C}{2}(\frac{1}{\Gamma}-\cos{2\alpha}\frac{\Gamma}{\Gamma^{2}+4w^{2}_{L}}\nonumber
      -\sin2\alpha\frac{2w_{L}}{\Gamma^{2}+4w^{2}_{L}}).\nonumber\\
\end{eqnarray}

where $I_{0}$ is the initial intensity and C is the constant for the fluorescence collection. $\Gamma$ is the spontaneous radiation linewidth of the excited state. For the Ca 423 nm transition, $\Gamma=2\pi\times34.6$ MHz.  $w_{L}=g_{J}(\mu_{B}/\hbar)B$ is the Larmor frequency and $\mu_{B}=e\hbar/2m_{e}$ is the Bohr magneton. For observing the Hanle effect, we added a polarizer in front of the PMT. Through a polarizer, $\alpha$ is the angle between the transmission axis of the polarizer and the polarization orientation of the fluorescence.

Thus, the shape of the Hanle effect signal with the magnetic field being swept through zero depends on the orientation of the polarizer. As shown in Figure 4, the Hanle effect signals have a Lorentzian shape for $\alpha=0$ or $\frac{\pi}{2}$ and a dispersion shape for $\alpha=\frac{\pi}{4}$ or $\frac{3\pi}{4}$ while the magnetic field is swept though zero. The reason for the incomplete dispersion shape is that the increasing Zeeman splitting with the increasing magnetic field decreases the signal intensity within the finite Doppler broadened spectra.

\begin{figure}
\includegraphics[width=9cm]{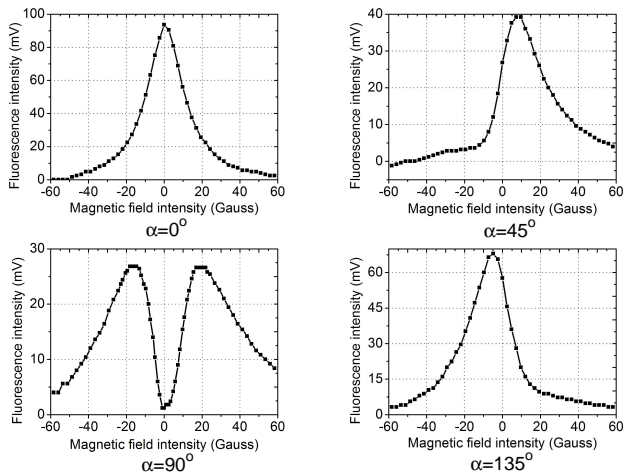}
\caption{\label{fig:epsart} The Hanle effect signal in different $\alpha$. The Lorentzian shape is observed for $\alpha=0$ and $\frac{\pi}{2}$. The incomplete dispersion shape is observed for $\alpha=\frac{\pi}{4}$ and $\frac{3\pi}{4}$.}
\end{figure}

\begin{figure}
\centering\includegraphics[width=9cm]{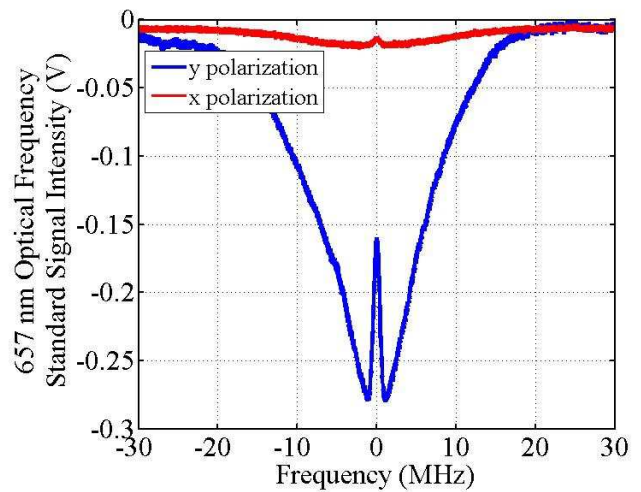}
\caption{\label{fig:epsart}The detected 657 nm optical frequency standard signal by using 423 nm electron-shelving method in Ca atomic beam optical frequency standard with and without optimized Hanle detection configuration. The 657 nm light is from the semiconductor laser.}
\end{figure}

With the fixed 423 nm fluorescence, we detect the 657 nm saturation spectroscopy signal by the "cat eye" structure. In Figure 5, the detected 657 nm saturation spectroscopy intensity with the $\hat{y}-$direction polarization of the 423 nm detection laser is 20.4 times larger than 657 nm saturation spectroscopy intensity with the $\hat{x}-$direction polarization of the 423 nm detection laser, which will bring one order of magnitude improvement in the performance of Ca atomic beam optical frequency standard.

Based on the Hanle effect in Ca atomic beam optical frequency standard, the most effective fluorescence collection geometry as shown in Figure 2 is that the 423 nm detection laser should be polarized in $\hat{y}-$direction while the magnetic field and the PMT is in the $\hat{x}-$direction. An optimized polarization orientation of the detection laser should be chosen for a fixed magnetic field direction and detector position, or an optimized detector position should be chosen for a fixed magnetic field direction and polarization orientation of the detection laser in a settled optical clock. The basic principle of the Hanle detection configuration is to set the PMT at the plane which is perpendicular to the polarization direction of the detection laser which is parallel to the magnetic field direction, or at the position along the magnetic field direction while the magnetic field direction is perpendicular to the polarization direction of the detection laser. Besides, we design a fluorescence collection structure for increasing the solid angle of the fluorescence collection in the compact Ca atomic beam optical frequency standard.

In Figure 2, the cylinder fluorescence collection structure with two through-holes in the dashed box is added. In the bottom of the cylinder, there is a 95\% reflection concave mirror. Also, in the top of the cylinder, there is a 95\% reflection concave mirror with a hole for the output of the fluorescence to PMT. The two reflection mirrors increase the solid angle of the fluorescence collection and average out spatial fluorescence distribution of the Hanle effect. Comparing with the Figure 3, the results show that the detected fluorescence intensity with the $\hat{y}-$direction polarization of the detection laser is only 1.3 times larger than the detected fluorescence intensity with the $\hat{x}-$direction polarization of the detection laser. The influence of the Hanle effect is averaged.

 Currently, the best performance of optical lattice clocks has reached $10^{-18}$ stability and uncertainty\cite{Lattice1, Ye}. In order to further improve the performance of the optical lattice clock, the improved fluorescence collection rate will be helpful\cite{Lattice1, Ye, Lattice3, Ch}. A normal fluorescence collection configuration of electron-shelving method in optical lattice clock\cite{Lattice3} is shown in Figure 6, where two counter-propagation detection lasers are circular polarization and in opposite polarization orientation. In this case, the detection laser can be taken as a linear polarization laser and rotates around the propagation of the laser. The maximum fluorescence intensity is in the propagation direction of the detection laser. When the PMT is set to the position in Figure 6, only half of the fluorescence intensity is detected. If we use an independent linearly polarization detection laser in optical lattice clocks\cite{Lattice1, Ye, Lattice3, Ch}, the most efficient Hanle detection configuration is to set the PMT at the plane which is perpendicular to the polarization direction of the detection laser which is parallel to the magnetic field direction, or at the position along the magnetic field direction while the magnetic field direction is perpendicular to the polarization direction of the detection laser. Or the second choice is the previous configuration with the PMT in the propagation direction of the detection laser. Otherwise, the fluorescence collection rate will decrease by more than one order of magnitude with the inefficient geometry. Our experimental results demonstrate the optimized Hanle detection geometry can improve the fluorescence collection rate for the optical lattice clock and further improve the performance of the optical lattice clock.

\begin{figure}
\includegraphics[width=8cm]{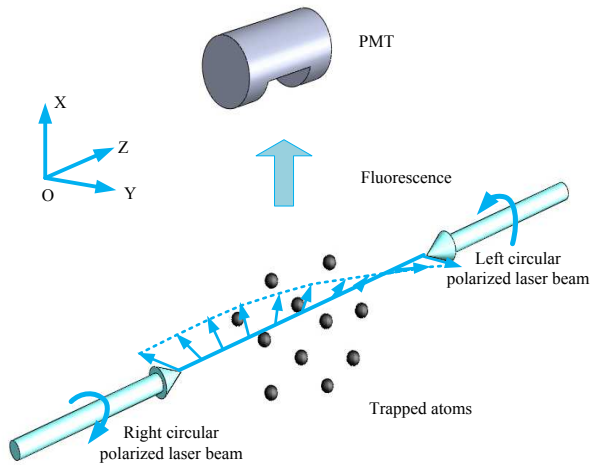}
\caption{\label{fig:epsart}A normal fluorescence collection configuration of electron-shelving method in Sr optical lattice clock.}
\end{figure}

 In conclusion, due to the Hanle effect, the fluorescence collection rate which changes 15.8 times is observed in the compact Ca atomic beam optical frequency standard. We thus propose the Hanle detection configuration of the electron-shelving method for optical clocks with limited solid angle of the fluorescence collection. The principle of the Hanle detection is to set the PMT at the plane which is perpendicular to the polarization direction of the detection laser which is parallel to the magnetic field direction, or at the position along the magnetic field direction while the magnetic field direction is perpendicular to the polarization direction of the detection laser, which is the most efficient Hanle detection configuration. Also, one can use the $\delta^{\pm}$ polarization detection laser in counter-propagation direction with the PMT in the propagation direction of the detection laser. With the optimized Hanle detection configuration, the 657 nm optical frequency standard signal intensity is improved 20 times, which will enhance the performance of the Ca atomic beam optical frequency standard. It is effective for the optical lattice clock, ion clock and quantum information experiments\cite{Lattice1, Ye, Lattice3, Ch, Gao, Qu, Quantum}. The optimized Hanle detection configuration will be helpful to further improve the performance of optical clocks\cite{Lattice1, Ye, Lattice3, Ch}. Besides, we design a fluorescence collection structure to improve the solid angle of the fluorescence collection in the compact Ca atomic beam optical frequency standard and it averages the influence of the Hanle effect in the fluorescence collection.

This work was supported by the National Natural Science Foundation of China (Grant Nos.10874009 and 11074011).

\bibliography{zxg}

\begin{thebibliography}{24}%
\makeatletter
\providecommand \@ifxundefined [1]{%
 \@ifx{#1\undefined}
}%
\providecommand \@ifnum [1]{%
 \ifnum #1\expandafter \@firstoftwo
 \else \expandafter \@secondoftwo
 \fi
}%
\providecommand \@ifx [1]{%
 \ifx #1\expandafter \@firstoftwo
 \else \expandafter \@secondoftwo
 \fi
}%
\providecommand \natexlab [1]{#1}%
\providecommand \enquote  [1]{``#1''}%
\providecommand \bibnamefont  [1]{#1}%
\providecommand \bibfnamefont [1]{#1}%
\providecommand \citenamefont [1]{#1}%
\providecommand \href@noop [0]{\@secondoftwo}%
\providecommand \href [0]{\begingroup \@sanitize@url \@href}%
\providecommand \@href[1]{\@@startlink{#1}\@@href}%
\providecommand \@@href[1]{\endgroup#1\@@endlink}%
\providecommand \@sanitize@url [0]{\catcode `\\12\catcode `\$12\catcode
  `\&12\catcode `\#12\catcode `\^12\catcode `\_12\catcode `\%12\relax}%
\providecommand \@@startlink[1]{}%
\providecommand \@@endlink[0]{}%
\providecommand \url  [0]{\begingroup\@sanitize@url \@url }%
\providecommand \@url [1]{\endgroup\@href {#1}{\urlprefix }}%
\providecommand \urlprefix  [0]{URL }%
\providecommand \Eprint [0]{\href }%
\providecommand \doibase [0]{http://dx.doi.org/}%
\providecommand \selectlanguage [0]{\@gobble}%
\providecommand \bibinfo  [0]{\@secondoftwo}%
\providecommand \bibfield  [0]{\@secondoftwo}%
\providecommand \translation [1]{[#1]}%
\providecommand \BibitemOpen [0]{}%
\providecommand \bibitemStop [0]{}%
\providecommand \bibitemNoStop [0]{.\EOS\space}%
\providecommand \EOS [0]{\spacefactor3000\relax}%
\providecommand \BibitemShut  [1]{\csname bibitem#1\endcsname}%
\let\auto@bib@innerbib\@empty
\bibitem [{\citenamefont {Hanle}(1923)}]{Hanle1923}%
  \BibitemOpen
  \bibfield  {author} {\bibinfo {author} {\bibfnamefont {W.}~\bibnamefont
  {Hanle}},\ }\href@noop {} {\bibfield  {journal} {\bibinfo  {journal}
  {Naturwissenschaften}\ }\textbf {\bibinfo {volume} {11}},\ \bibinfo {pages}
  {690--691} (\bibinfo {year} {1923})}\BibitemShut {NoStop}%
\bibitem [{\citenamefont {Hanle}(1924)}]{Hanle1924}%
  \BibitemOpen
  \bibfield  {author} {\bibinfo {author} {\bibfnamefont {W.}~\bibnamefont
  {Hanle}},\ }\href@noop {} {\bibfield  {journal} {\bibinfo  {journal} {Z.
  Phys.}\ }\textbf {\bibinfo {volume} {30}},\ \bibinfo {pages} {93--105}
  (\bibinfo {year} {1924})}\BibitemShut {NoStop}%
\bibitem [{\citenamefont {de~Zafra}\ and\ \citenamefont
  {Kirk}(1967)}]{lifetime1}%
  \BibitemOpen
  \bibfield  {author} {\bibinfo {author} {\bibfnamefont {R.}~\bibnamefont
  {de~Zafra}}\ and\ \bibinfo {author} {\bibfnamefont {W.}~\bibnamefont
  {Kirk}},\ }\href@noop {} {\bibfield  {journal} {\bibinfo  {journal} {Amer. J.
  Phys.}\ }\textbf {\bibinfo {volume} {35}},\ \bibinfo {pages} {573} (\bibinfo
  {year} {1967})}\BibitemShut {NoStop}%
\bibitem [{\citenamefont {Lurio}, \citenamefont {Dezafra},\ and\ \citenamefont
  {Goshen}(1964)}]{lifetime2}%
  \BibitemOpen
  \bibfield  {author} {\bibinfo {author} {\bibfnamefont {A.}~\bibnamefont
  {Lurio}}, \bibinfo {author} {\bibfnamefont {R.~L.}\ \bibnamefont {Dezafra}},
  \ and\ \bibinfo {author} {\bibfnamefont {R.}~\bibnamefont {Goshen}},\
  }\href@noop {} {\bibfield  {journal} {\bibinfo  {journal} {Phys. R}\ }\textbf
  {\bibinfo {volume} {134}},\ \bibinfo {pages} {1198--1203} (\bibinfo {year}
  {1964})}\BibitemShut {NoStop}%
\bibitem [{\citenamefont {Kelly}\ and\ \citenamefont
  {Mathur}(1980)}]{lifetime3}%
  \BibitemOpen
  \bibfield  {author} {\bibinfo {author} {\bibfnamefont {F.~M.}\ \bibnamefont
  {Kelly}}\ and\ \bibinfo {author} {\bibfnamefont {M.~S.}\ \bibnamefont
  {Mathur}},\ }\href@noop {} {\bibfield  {journal} {\bibinfo  {journal} {Can.
  J. Phys.}\ }\textbf {\bibinfo {volume} {58}},\ \bibinfo {pages} {1416}
  (\bibinfo {year} {1980})}\BibitemShut {NoStop}%
\bibitem [{\citenamefont {Moruzzi}\ and\ \citenamefont
  {Strumia}(1991)}]{Hanle}%
  \BibitemOpen
  \bibfield  {author} {\bibinfo {author} {\bibfnamefont {G.}~\bibnamefont
  {Moruzzi}}\ and\ \bibinfo {author} {\bibfnamefont {F.}~\bibnamefont
  {Strumia}},\ }\href@noop {} {\emph {\bibinfo {title} {The Hanle effect and
  level-crossing spectroscopy}}}\ (\bibinfo  {publisher} {Springer Science $+$
  Business Media New York},\ \bibinfo {year} {1991})\BibitemShut {NoStop}%
\bibitem [{\citenamefont {Franken}(1961)}]{atomic5}%
  \BibitemOpen
  \bibfield  {author} {\bibinfo {author} {\bibfnamefont {P.~A.}\ \bibnamefont
  {Franken}},\ }\href@noop {} {\bibfield  {journal} {\bibinfo  {journal}
  {Physical Review}\ }\textbf {\bibinfo {volume} {121}},\ \bibinfo {pages}
  {508--513} (\bibinfo {year} {1961})}\BibitemShut {NoStop}%
\bibitem [{\citenamefont {Jun}\ and\ \citenamefont {Lee}(1998)}]{atomic1}%
  \BibitemOpen
  \bibfield  {author} {\bibinfo {author} {\bibfnamefont {J.~W.}\ \bibnamefont
  {Jun}}\ and\ \bibinfo {author} {\bibfnamefont {H.~S.}\ \bibnamefont {Lee}},\
  }\href@noop {} {\bibfield  {journal} {\bibinfo  {journal} {Optics
  Communication}\ }\textbf {\bibinfo {volume} {149}},\ \bibinfo {pages}
  {43--49} (\bibinfo {year} {1998})}\BibitemShut {NoStop}%
\bibitem [{\citenamefont {Kolwas}(1977)}]{atomic2}%
  \BibitemOpen
  \bibfield  {author} {\bibinfo {author} {\bibfnamefont {M.}~\bibnamefont
  {Kolwas}},\ }\href@noop {} {\bibfield  {journal} {\bibinfo  {journal} {J.
  Phys. B}\ }\textbf {\bibinfo {volume} {10}},\ \bibinfo {pages} {583--594}
  (\bibinfo {year} {1977})}\BibitemShut {NoStop}%
\bibitem [{\citenamefont {Yu}\ \emph {et~al.}(2012)\citenamefont {Yu},
  \citenamefont {Zhong}, \citenamefont {Wang},\ and\ \citenamefont
  {Zhan}}]{Yu}%
  \BibitemOpen
  \bibfield  {author} {\bibinfo {author} {\bibfnamefont {G.}~\bibnamefont
  {Yu}}, \bibinfo {author} {\bibfnamefont {J.}~\bibnamefont {Zhong}}, \bibinfo
  {author} {\bibfnamefont {J.}~\bibnamefont {Wang}}, \ and\ \bibinfo {author}
  {\bibfnamefont {M.}~\bibnamefont {Zhan}},\ }\href@noop {} {\bibfield
  {journal} {\bibinfo  {journal} {Chin. J. Quant. Electr.}\ }\textbf {\bibinfo
  {volume} {29}},\ \bibinfo {pages} {257--264} (\bibinfo {year}
  {2012})}\BibitemShut {NoStop}%
\bibitem [{\citenamefont {Labeyrie}\ \emph {et~al.}(2002)\citenamefont
  {Labeyrie}, \citenamefont {Miniatura}, \citenamefont {Mueller}, \citenamefont
  {Sigwarth}, \citenamefont {Delande},\ and\ \citenamefont {Kaiser}}]{atomic6}%
  \BibitemOpen
  \bibfield  {author} {\bibinfo {author} {\bibfnamefont {G.}~\bibnamefont
  {Labeyrie}}, \bibinfo {author} {\bibfnamefont {C.}~\bibnamefont {Miniatura}},
  \bibinfo {author} {\bibfnamefont {C.~A.}\ \bibnamefont {Mueller}}, \bibinfo
  {author} {\bibfnamefont {O.}~\bibnamefont {Sigwarth}}, \bibinfo {author}
  {\bibfnamefont {D.}~\bibnamefont {Delande}}, \ and\ \bibinfo {author}
  {\bibfnamefont {R.}~\bibnamefont {Kaiser}},\ }\href@noop {} {\bibfield
  {journal} {\bibinfo  {journal} {Phys. Rev. Lett.}\ }\textbf {\bibinfo
  {volume} {89}},\ \bibinfo {pages} {163901} (\bibinfo {year}
  {2002})}\BibitemShut {NoStop}%
\bibitem [{\citenamefont {Auzinsh}\ \emph {et~al.}(2013)\citenamefont
  {Auzinsh}, \citenamefont {A.Berzins}, \citenamefont {Ferber}, \citenamefont
  {Gahbauer}, \citenamefont {Kalvans}, \citenamefont {Mozers},\ and\
  \citenamefont {Spiss}}]{atomic3}%
  \BibitemOpen
  \bibfield  {author} {\bibinfo {author} {\bibfnamefont {M.}~\bibnamefont
  {Auzinsh}}, \bibinfo {author} {\bibnamefont {A.Berzins}}, \bibinfo {author}
  {\bibfnamefont {R.}~\bibnamefont {Ferber}}, \bibinfo {author} {\bibfnamefont
  {F.}~\bibnamefont {Gahbauer}}, \bibinfo {author} {\bibfnamefont
  {L.}~\bibnamefont {Kalvans}}, \bibinfo {author} {\bibfnamefont
  {A.}~\bibnamefont {Mozers}}, \ and\ \bibinfo {author} {\bibfnamefont
  {A.}~\bibnamefont {Spiss}},\ }\href@noop {} {\bibfield  {journal} {\bibinfo
  {journal} {Phys. Rev. A}\ }\textbf {\bibinfo {volume} {87}},\ \bibinfo
  {pages} {033412} (\bibinfo {year} {2013})}\BibitemShut {NoStop}%
\bibitem [{\citenamefont {Avan}\ and\ \citenamefont
  {Cohen-Tannoudji}(1977)}]{Co}%
  \BibitemOpen
  \bibfield  {author} {\bibinfo {author} {\bibfnamefont {P.}~\bibnamefont
  {Avan}}\ and\ \bibinfo {author} {\bibfnamefont {C.}~\bibnamefont
  {Cohen-Tannoudji}},\ }\href@noop {} {\bibfield  {journal} {\bibinfo
  {journal} {J. Phys. B: Atom. Molec. Phys.}\ }\textbf {\bibinfo {volume}
  {10}},\ \bibinfo {pages} {171--185} (\bibinfo {year} {1977})}\BibitemShut
  {NoStop}%
\bibitem [{\citenamefont {c$\acute{e}$rez}\ \emph {et~al.}(1991)\citenamefont
  {c$\acute{e}$rez}, \citenamefont {Th$\acute{e}$obald}, \citenamefont
  {Giordano}, \citenamefont {Dimarcq},\ and\ \citenamefont
  {Labachelevie}}]{Cs1}%
  \BibitemOpen
  \bibfield  {author} {\bibinfo {author} {\bibfnamefont {P.}~\bibnamefont
  {c$\acute{e}$rez}}, \bibinfo {author} {\bibfnamefont {G.}~\bibnamefont
  {Th$\acute{e}$obald}}, \bibinfo {author} {\bibfnamefont {V.}~\bibnamefont
  {Giordano}}, \bibinfo {author} {\bibfnamefont {N.}~\bibnamefont {Dimarcq}}, \
  and\ \bibinfo {author} {\bibfnamefont {M.}~\bibnamefont {Labachelevie}},\
  }\href@noop {} {\bibfield  {journal} {\bibinfo  {journal} {IEEE Transactions
  on Instrumentation and Measurement}\ }\textbf {\bibinfo {volume} {40}},\
  \bibinfo {pages} {137--141} (\bibinfo {year} {1991})}\BibitemShut {NoStop}%
\bibitem [{\citenamefont {Th$\acute{e}$obald}\ \emph
  {et~al.}(1989)\citenamefont {Th$\acute{e}$obald}, \citenamefont {Dimarcq},
  \citenamefont {Giordano},\ and\ \citenamefont {c$\acute{e}$rez}}]{Cs2}%
  \BibitemOpen
  \bibfield  {author} {\bibinfo {author} {\bibfnamefont {G.}~\bibnamefont
  {Th$\acute{e}$obald}}, \bibinfo {author} {\bibfnamefont {N.}~\bibnamefont
  {Dimarcq}}, \bibinfo {author} {\bibfnamefont {V.}~\bibnamefont {Giordano}}, \
  and\ \bibinfo {author} {\bibfnamefont {P.}~\bibnamefont {c$\acute{e}$rez}},\
  }\href@noop {} {\bibfield  {journal} {\bibinfo  {journal} {Optics
  Communications}\ }\textbf {\bibinfo {volume} {71}},\ \bibinfo {pages}
  {256--262} (\bibinfo {year} {1989})}\BibitemShut {NoStop}%
\bibitem [{\citenamefont {Huang}\ \emph {et~al.}(2006)\citenamefont {Huang},
  \citenamefont {Zhang}, \citenamefont {Yu}, \citenamefont {Chen},
  \citenamefont {Zhuang},\ and\ \citenamefont {Chen}}]{Ca-p}%
  \BibitemOpen
  \bibfield  {author} {\bibinfo {author} {\bibfnamefont {K.}~\bibnamefont
  {Huang}}, \bibinfo {author} {\bibfnamefont {J.}~\bibnamefont {Zhang}},
  \bibinfo {author} {\bibfnamefont {D.}~\bibnamefont {Yu}}, \bibinfo {author}
  {\bibfnamefont {Z.}~\bibnamefont {Chen}}, \bibinfo {author} {\bibfnamefont
  {W.}~\bibnamefont {Zhuang}}, \ and\ \bibinfo {author} {\bibfnamefont
  {J.}~\bibnamefont {Chen}},\ }\href@noop {} {\bibfield  {journal} {\bibinfo
  {journal} {Chin. Phys. Lett.}\ }\textbf {\bibinfo {volume} {23}},\ \bibinfo
  {pages} {3198--3201} (\bibinfo {year} {2006})}\BibitemShut {NoStop}%
\bibitem [{\citenamefont {Hinkley}\ \emph {et~al.}(2013)\citenamefont
  {Hinkley}, \citenamefont {Sherman}, \citenamefont {Phillips}, \citenamefont
  {Schioppo}, \citenamefont {Lemke}, \citenamefont {Beloy}, \citenamefont
  {Pizzocaro}, \citenamefont {Oates},\ and\ \citenamefont {Ludlow}}]{Lattice1}%
  \BibitemOpen
  \bibfield  {author} {\bibinfo {author} {\bibfnamefont {N.}~\bibnamefont
  {Hinkley}}, \bibinfo {author} {\bibfnamefont {J.~A.}\ \bibnamefont
  {Sherman}}, \bibinfo {author} {\bibfnamefont {N.~B.}\ \bibnamefont
  {Phillips}}, \bibinfo {author} {\bibfnamefont {M.}~\bibnamefont {Schioppo}},
  \bibinfo {author} {\bibfnamefont {N.~D.}\ \bibnamefont {Lemke}}, \bibinfo
  {author} {\bibfnamefont {K.}~\bibnamefont {Beloy}}, \bibinfo {author}
  {\bibfnamefont {M.}~\bibnamefont {Pizzocaro}}, \bibinfo {author}
  {\bibfnamefont {C.~W.}\ \bibnamefont {Oates}}, \ and\ \bibinfo {author}
  {\bibfnamefont {A.~D.}\ \bibnamefont {Ludlow}},\ }\href@noop {} {\bibfield
  {journal} {\bibinfo  {journal} {Science}\ }\textbf {\bibinfo {volume}
  {341}},\ \bibinfo {pages} {1215--1218} (\bibinfo {year} {2013})}\BibitemShut
  {NoStop}%
\bibitem [{\citenamefont {Bloom}\ \emph {et~al.}(2013)\citenamefont {Bloom},
  \citenamefont {Nicholson}, \citenamefont {Williams}, \citenamefont
  {Compbell}, \citenamefont {Bishof}, \citenamefont {Zhang}, \citenamefont
  {Zhang}, \citenamefont {Bromley},\ and\ \citenamefont {Ye}}]{Ye}%
  \BibitemOpen
  \bibfield  {author} {\bibinfo {author} {\bibfnamefont {B.~J.}\ \bibnamefont
  {Bloom}}, \bibinfo {author} {\bibfnamefont {T.~L.}\ \bibnamefont
  {Nicholson}}, \bibinfo {author} {\bibfnamefont {J.~R.}\ \bibnamefont
  {Williams}}, \bibinfo {author} {\bibfnamefont {S.~L.}\ \bibnamefont
  {Compbell}}, \bibinfo {author} {\bibfnamefont {M.}~\bibnamefont {Bishof}},
  \bibinfo {author} {\bibfnamefont {X.}~\bibnamefont {Zhang}}, \bibinfo
  {author} {\bibfnamefont {W.}~\bibnamefont {Zhang}}, \bibinfo {author}
  {\bibfnamefont {S.~L.}\ \bibnamefont {Bromley}}, \ and\ \bibinfo {author}
  {\bibfnamefont {J.}~\bibnamefont {Ye}},\ }\href@noop {} {\bibfield  {journal}
  {\bibinfo  {journal} {ArXiv: physics/1309.1107}\ } (\bibinfo {year}
  {2013})}\BibitemShut {NoStop}%
\bibitem [{\citenamefont {Wang}\ \emph {et~al.}(2009)\citenamefont {Wang},
  \citenamefont {Wang}, \citenamefont {Lin}, \citenamefont {Wang},
  \citenamefont {Lin}, \citenamefont {Zang}, \citenamefont {Li},\ and\
  \citenamefont {Fang}}]{Lattice3}%
  \BibitemOpen
  \bibfield  {author} {\bibinfo {author} {\bibfnamefont {S.}~\bibnamefont
  {Wang}}, \bibinfo {author} {\bibfnamefont {Q.}~\bibnamefont {Wang}}, \bibinfo
  {author} {\bibfnamefont {Y.}~\bibnamefont {Lin}}, \bibinfo {author}
  {\bibfnamefont {M.}~\bibnamefont {Wang}}, \bibinfo {author} {\bibfnamefont
  {B.}~\bibnamefont {Lin}}, \bibinfo {author} {\bibfnamefont {E.}~\bibnamefont
  {Zang}}, \bibinfo {author} {\bibfnamefont {T.}~\bibnamefont {Li}}, \ and\
  \bibinfo {author} {\bibfnamefont {Z.}~\bibnamefont {Fang}},\ }\href@noop {}
  {\bibfield  {journal} {\bibinfo  {journal} {Chin. Phys. Lett.}\ }\textbf
  {\bibinfo {volume} {26}},\ \bibinfo {pages} {093202} (\bibinfo {year}
  {2009})}\BibitemShut {NoStop}%
\bibitem [{\citenamefont {Chen}\ \emph {et~al.}(2013)\citenamefont {Chen},
  \citenamefont {Zhou}, \citenamefont {Chen}, \citenamefont {Fang},
  \citenamefont {Huang}, \citenamefont {Zhang}, \citenamefont {Gao},
  \citenamefont {Jiang}, \citenamefont {Bi}, \citenamefont {Ma},\ and\
  \citenamefont {Xu}}]{Ch}%
  \BibitemOpen
  \bibfield  {author} {\bibinfo {author} {\bibfnamefont {N.}~\bibnamefont
  {Chen}}, \bibinfo {author} {\bibfnamefont {M.}~\bibnamefont {Zhou}}, \bibinfo
  {author} {\bibfnamefont {H.}~\bibnamefont {Chen}}, \bibinfo {author}
  {\bibfnamefont {S.}~\bibnamefont {Fang}}, \bibinfo {author} {\bibfnamefont
  {L.}~\bibnamefont {Huang}}, \bibinfo {author} {\bibfnamefont
  {X.}~\bibnamefont {Zhang}}, \bibinfo {author} {\bibfnamefont
  {Q.}~\bibnamefont {Gao}}, \bibinfo {author} {\bibfnamefont {Y.}~\bibnamefont
  {Jiang}}, \bibinfo {author} {\bibfnamefont {Z.}~\bibnamefont {Bi}}, \bibinfo
  {author} {\bibfnamefont {L.}~\bibnamefont {Ma}}, \ and\ \bibinfo {author}
  {\bibfnamefont {X.}~\bibnamefont {Xu}},\ }\href@noop {} {\bibfield  {journal}
  {\bibinfo  {journal} {Chin. Phys. B}\ }\textbf {\bibinfo {volume} {22}},\
  \bibinfo {pages} {090601} (\bibinfo {year} {2013})}\BibitemShut {NoStop}%
\bibitem [{\citenamefont {Gao}(2013)}]{Gao}%
  \BibitemOpen
  \bibfield  {author} {\bibinfo {author} {\bibfnamefont {K.}~\bibnamefont
  {Gao}},\ }\href@noop {} {\bibfield  {journal} {\bibinfo  {journal} {Chinese
  Science Bulletin}\ }\textbf {\bibinfo {volume} {58}},\ \bibinfo {pages}
  {853--863} (\bibinfo {year} {2013})}\BibitemShut {NoStop}%
\bibitem [{\citenamefont {McFerran}\ and\ \citenamefont {Luiten}(2010)}]{Ca-a}%
  \BibitemOpen
  \bibfield  {author} {\bibinfo {author} {\bibfnamefont {J.~J.}\ \bibnamefont
  {McFerran}}\ and\ \bibinfo {author} {\bibfnamefont {A.~N.}\ \bibnamefont
  {Luiten}},\ }\href@noop {} {\bibfield  {journal} {\bibinfo  {journal} {J.
  Opt. Soc. Am. B}\ }\textbf {\bibinfo {volume} {27}},\ \bibinfo {pages}
  {277--285} (\bibinfo {year} {2010})}\BibitemShut {NoStop}%
\bibitem [{\citenamefont {Pruttivarasin}, \citenamefont {Ramm},\ and\
  \citenamefont {H$\ddot{a}$ffner}(2013)}]{Qu}%
  \BibitemOpen
  \bibfield  {author} {\bibinfo {author} {\bibfnamefont {T.}~\bibnamefont
  {Pruttivarasin}}, \bibinfo {author} {\bibfnamefont {M.}~\bibnamefont {Ramm}},
  \ and\ \bibinfo {author} {\bibfnamefont {H.}~\bibnamefont
  {H$\ddot{a}$ffner}},\ }\href@noop {} {\bibfield  {journal} {\bibinfo
  {journal} {ArXiv: physics. atom-ph/1312.7617}\ } (\bibinfo {year}
  {2013})}\BibitemShut {NoStop}%
\bibitem [{\citenamefont {Benhelm}\ \emph {et~al.}(2008)\citenamefont
  {Benhelm}, \citenamefont {Kirchmair}, \citenamefont {Roos},\ and\
  \citenamefont {Blatt}}]{Quantum}%
  \BibitemOpen
  \bibfield  {author} {\bibinfo {author} {\bibfnamefont {J.}~\bibnamefont
  {Benhelm}}, \bibinfo {author} {\bibfnamefont {G.}~\bibnamefont {Kirchmair}},
  \bibinfo {author} {\bibfnamefont {C.~F.}\ \bibnamefont {Roos}}, \ and\
  \bibinfo {author} {\bibfnamefont {R.}~\bibnamefont {Blatt}},\ }\href@noop {}
  {\bibfield  {journal} {\bibinfo  {journal} {Phys. Rev. A}\ }\textbf {\bibinfo
  {volume} {77}},\ \bibinfo {pages} {062306} (\bibinfo {year}
  {2008})}\BibitemShut {NoStop}%
\end{thebibliography}%

\end{document}